\documentclass[prd,aps,preprintnumbers,fleqn,showpacs,nofootinbib,superscriptaddress]{revtex4}

\usepackage{amssymb,amsmath,graphicx,bm}

\usepackage[usenames,dvipsnames]{color}
\usepackage[normalem]{ulem} % \sout{old text} for strikeout
\renewcommand\sout{\bgroup \color[rgb]{0.55,0.00,0.99} \ULdepth=-.5ex \ULset}

\def\slash#1{\setbox0=\hbox{$#1$}               % set a box for #1
        \dimen0=\wd0                            % and get its size
        \setbox1=\hbox{/} \dimen1=\wd1          % get size of /
        \ifdim\dimen0>\dimen1                   % #1 is bigger
        \rlap{\hbox to \dimen0{\hfil/\hfil}}    % so center / in box
        #1                                      % and print #1
        \else
        \rlap{\hbox to \dimen1{\hfil$#1$\hfil}} % so center #1
        /                                       % and print /
        \fi}                                    %

\newcommand{\beqn}{\begin{eqnarray}}
\newcommand{\eeqn}{\end{eqnarray}}

\begin{document}

\title{Probing the linear polarization of photons in ultraperipheral heavy ion collisions}

\author{Cong Li}
 \affiliation{\normalsize\it Key Laboratory of
Particle Physics and Particle Irradiation (MOE),Institute of
frontier and interdisciplinary science, Shandong University,
QingDao, Shandong 266237, China }

\author{Jian~Zhou}
 \affiliation{\normalsize\it Key Laboratory of
Particle Physics and Particle Irradiation (MOE),Institute of
frontier and interdisciplinary science, Shandong University,
QingDao, Shandong 266237, China }

\author{Ya-jin Zhou}
\affiliation{\normalsize\it Key Laboratory of Particle Physics and
Particle Irradiation (MOE),Institute of frontier and
interdisciplinary science, Shandong University, QingDao, Shandong
266237, China }

\begin{abstract}
We propose to measure the linear polarization of
 the external electromagnetic fields of a  relativistic  heavy ion through azimuthal asymmetries
in dilepton production in ultraperipheral collisions. The
asymmetries estimated with the equivalent photon approximation are
 shown to be sizable.
\end{abstract}

% \pacs{...}
\date{\today}

\maketitle

\section{Introduction}
Transverse momentum dependent(TMD) parton distribution function~\cite{Collins:1981uk} is one of the most powerful theoretical
 tools that are utilized to explore the three-dimensional imaging of nuleon/nuclei. Among many TMD parton
 distributions, the linearly polarized gluon distribution~\cite{Mulders:2000sh} has received growing attentions in
 recent years. It describes the correlation between gluon transverse momentum and its polarization vector inside
 an unpolarized nucleon or nucleus. It is of particular interest to study linearly polarized gluon
 distribution at small $x$~\cite{Metz:2011wb,Dominguez:2011br}, as
 it is predicted to grow equally rapidly towards small $x$ as compared to the unpolarized gluon distribution in the
 dilute limit.  In the saturation  limit, the dipole type linearly polarized gluon distribution
  and the dipole type unpolarized gluon distribution remain identical, whereas the
  linearly polarization of Weizs$\ddot{a}$cker-Williams gluons is suppressed.  Though it has been found
  promising to probe the linearly polarized gluon distribution by measuring $\cos 2\phi$ azimuthal asymmetry
  for two particle production in various high energy scattering processes at RHIC, LHC, or
  a future Electron-Ion Collider(EIC)~\cite{Metz:2011wb,Dominguez:2011br,Boer:2009nc,Boer:2010zf,Qiu:2011ai,Pisano:2013cya,Akcakaya:2012si,Dumitru:2015gaa,Kotko:2015ura,Boer:2017xpy,Marquet:2017xwy},
  this gluon distribution so far has not yet been studied experimentally.

 In  analogy to the QCD case, one  also can define a linearly polarized photon distribution for an
 unpolarized nucleon or nuclei target, which can be accessed by measuring the azimuthal asymmetries in
 di-lepton production  in hadron-hadron collisions~\cite{Pisano:2013cya}.  However,  it is not very practical  to extract
 the polarized photon distribution in hadronic reactions due to the di-lepton Drell-Yan production background.
 Instead, the  cleaner  and more promising way to probe the linearly polarization of photons
 would be the purely electromagnetic two photon reaction $\gamma \gamma \rightarrow l^+ l^-$
  in heavy-ion ultra-peripheral collisions(UPCs) where the hadronic background is absent.
  Though photon-photon collisions in the UPC case has been extensively studied
  \cite{Bertulani:1987tz,Vidovic:1992ik,Bertulani:2005ru,Baur:2007fv,Baltz:2007kq,Klein:2016yzr,Klein:2018cjh,Zha:2018ywo,Klein:2018fmp,Zha:2018tlq,Ye:2018jwq,Adams:2004rz,Aaboud:2018eph,Adam:2018tdm},
    to the best of our knowledge,
   the polarization dependent  effects have not yet been addressed so far.
 Both the  unpolarized photon distribution and the polarized one in the UPC case can
 be determined using the external classical field approximation~\cite{Bertulani:1987tz,Vidovic:1992ik}.
 It is not surprising to find that they are identical to each other in this approximation,
 just like the relation established between
  the dipole amplitude and the polarized gluon distributions~\cite{Metz:2011wb,Zhou:2013gsa,Boer:2015pni,Boer:2016xqr}.
  In the present paper, we propose to test this theoretical predication
  by measuring $\cos 2\phi$ and $\cos 4\phi $ asymmetries in di-lepton production induced by
  the linearly polarized photon distribution.

  Recently, the STAR collaboration at RHIC~\cite{Adam:2018tdm} and the ATLAS collaboration~\cite{Aaboud:2018eph} at LHC have carried out the
  measurements of transverse momentum spectra of lepton pairs for various invariant mass regions with
   high precision. The significant $q_\perp$(total transverse momentum of lepton pair) broadening effect found in hadronic heavy-ion collisions
  in comparison  to those in UPCs has stimulated a lot of theoretical
  progress~\cite{Klein:2018cjh,Zha:2018ywo,Klein:2018fmp,Zha:2018tlq,Ye:2018jwq}, as the
  transverse momentum broadening effect plays a crucial role in understanding the properties of the
  hot medium created in heavy-ion collisions.  Moreover, a small tail of events
  at high transverse momentum observed by the ATLAS offers a clean way to test the resummation
  formalism for the QED case~\cite{Klein:2018fmp}. Here we would like to point out that it is doable to extract linearly
  polarized photon distribution by analyzing the angular modulations of di-lepton production
  cross section from the existed experimental  data collected   by the STAR collaboration
  and the ATLAS collaboration. This analysis can be considered as a new way  to
  test how reliable the equivalent photon approximation widely used for computing  UPCs observables is.
 Furthermore, it  sets a baseline for studying the electromagnetic properties of QGP,
 since this contribution yields the asymmetries in hadronic  heavy-ion collisions as well.

The paper is structured as follows.  In the next section, we compute
 the azimuthal dependent cross section for the purely electromagnetic di-lepton production
 in terms of the linearly polarized photon distributions and the unpolarized photon distribution.
 We then present numerical results incorporating the Sudakov suppression effect for the asymmetries
 in the kinematical regions where the corresponding measurements have been carried out at RHIC and LHC.
  A summary of our findings and conclusions  is presented in Sec.III.

\section{Azimuthal asymmetries in di-lepton production in UPCs}
Di-lepton production in UPCs is well described by two photons reaction at the lowest order QED,
\begin{eqnarray}
\gamma(x_1P+k_{1\perp})+\gamma(x_2 \bar P+k_{2\perp}) \rightarrow l^+(p_1)+ l^-(p_2)
\end{eqnarray}
The leptons are produced nearly back-to-back in azimuthal with  total transverse momentum
$q_\perp\equiv p_{1\perp}+p_{2\perp}=k_{1\perp}+k_{2\perp}$ being much smaller than the individual
lepton transverse momenta $p_{1\perp}$ or $p_{2\perp}$. Since there are two well separated scales in this process,
 the application of TMD factorization is justified. If the calculation is formulated in TMD factorization,
 the two  leading power photon TMDs: the normal unpolarized photon TMD
 and the linearly polarized photon TMD contribute to the differential cross section. They are formally
 defined as the following,
\begin{eqnarray}
\int \frac{2dy^- d^2y_\perp}{xP^+(2\pi)^3} e^{ik \cdot y} \langle P
|  F_{+\perp}^\mu(0) F_{+\perp}^\nu(y)  |P \rangle \big|_{y^+=0}=
\delta_\perp^{\mu \nu} f_1^\gamma(x,k_\perp^2)+ \left (\frac{2k_\perp^\mu
k_\perp^\nu}{k_\perp^2}-\delta_\perp^{\mu\nu} \right )
 h_1^{\perp \gamma}(x,k_\perp^2) , \label{gmat}
\end{eqnarray}
where the transverse tensor is commonly defined:
 $\delta_\perp^{\mu\nu}=-g^{\mu\nu}+p^\mu n^\nu+p^\nu n^\mu$ and
$k_\perp^2=\delta_\perp^{\mu\nu} k_{\perp\mu} k_{\perp\nu}$.
Two photon TMDs, $f_1^\gamma$ and $h_1^{\perp \gamma}$, are the
unpolarized and linearly polarized photon distribution, respectively.
This matrix element definition for photon TMDs bears much resemblance to those for the gluon ones~\cite{Mulders:2000sh}.
However, one should note that there is no need to add  gauge link for ensuring gauge invariance
since photon does't carry charge. As such, the light cone singularity is absent for the photon TMD case.

One can easily recover the azimuthal dependent cross section for lepton pair production
from the results for heavy quark pair production existed in the literatures~\cite{Pisano:2013cya,Akcakaya:2012si}.
It is of course also straightforward to compute the cross section at the lowest order QED,
 which reads,
 \begin{eqnarray}
\frac{d\sigma}{d^2 p_{1\perp} d^2 p_{2\perp} dy_1 dy_2}= \frac{2\alpha_e^2}{Q^4}
\left [ \mathcal{A}+ \mathcal{B} \cos 2\phi+\mathcal{C} \cos 4\phi \right ]
\label{cs}
\end{eqnarray}
where $\phi$ is the angle between transverse momenta $q_\perp$ and
$P_\perp=(p_{1\perp}-p_{2\perp})/2$. $y_1$ and $y_2$ are leptons rapidities, respectively. Q is the
invariant mass of the lepton pair.
The coefficients $\mathcal{A}$, $\mathcal{B}$ and $\mathcal{C}$
contain convolutions of photon TMDs,
 \begin{eqnarray}
\mathcal{A}&=& \frac{(Q^2-2m^2)m^2+(Q^2-2 P_\perp^2)P_\perp^2}{(m^2+P_\perp^2)^2}
x_1x_2\!\int d^2k_{1\perp} d^2 k_{2\perp} \delta^2( q_\perp-k_{1\perp}-k_{2\perp})
f_1^\gamma(x_1,k_{1\perp}^2)f_1^\gamma(x_2,k_{2\perp}^2)
\nonumber \\ && +
\frac{m^4}{(m^2+P_\perp^2)^2} x_1x_2 \!
\int \!d^2k_{1\perp} d^2 k_{2\perp} \delta^2( q_\perp\!-k_{1\perp}\!-k_{2\perp})\!
\left [ 2(\hat k_{1\perp} \cdot \hat k_{2\perp})^2 \! -1 \right ]
h_1^{\perp \gamma}(x_1,k_{1\perp}^2)h_1^{\perp \gamma}(x_2,k_{2\perp}^2)
\end{eqnarray}
and
 \begin{eqnarray}
\mathcal{B}&=&
\frac{4m^2 P_{\perp}^2}{(m^2+P_\perp^2)^2}
x_1x_2\! \int \! d^2k_{1\perp} d^2 k_{2\perp} \delta^2( q_\perp\!-k_{1\perp}\!-k_{2\perp})
\nonumber \\&& \times  \left \{ \left [ 2(\hat k_{2\perp} \! \cdot \hat q_{\perp})^2\!-1\right ]
f_1^{ \gamma}\!(x_1,k_{1\perp}^2)h_1^{\perp \gamma}\!(x_2,k_{2\perp}^2) +
\left [ 2(\hat k_{1\perp} \!\cdot \hat q_{\perp})^2\!-1\right ]
h_1^{\perp \gamma}\!(x_1,k_{1\perp}^2)f_1^{\gamma}\!(x_2,k_{2\perp}^2)
\right \}
\end{eqnarray}
and
 \begin{eqnarray}
\mathcal{C} &=&
\frac{ -2P_{\perp}^4}{(m^2+P_\perp^2)^2}
x_1x_2\! \int \! d^2k_{1\perp} d^2 k_{2\perp} \delta^2( q_\perp\!-k_{1\perp}\!-k_{2\perp})
\nonumber \\
&&\times \left [2\left (\! 2(\hat k_{2\perp} \! \cdot \hat q_{\perp})(\hat k_{1\perp} \! \cdot \hat q_{\perp})
-\hat k_{1\perp} \!\cdot \!\hat k_{2\perp} \! \right )^2\!-1\right ]
 h_1^{ \perp \gamma}\!(x_1,k_{1\perp}^2) h_1^{\perp \gamma}\!(x_2,k_{2\perp}^2)
\end{eqnarray}
where $\hat k_{1\perp}$ and $\hat q_\perp$ are unit vectors defined as $\hat k_{1\perp}=k_{1\perp}/|{k_{1\perp}}|$
and $\hat q_\perp= q_\perp/| q_\perp| $ respectively.
The incoming photons  longitudinal momenta fraction  are fixed by the
external kinematics according to   $x_1=\sqrt{\frac{P_\perp^2+m^2}{s}}(e^{y_1}+e^{y_2})$
 and $x_2=\sqrt{\frac{P_\perp^2+m^2}{s}}(e^{-y_1}+e^{-y_2})$ with $m$ being lepton mass.

When going beyond the lowest order QED, the Sudakov type  logarithm terms
 $ \frac{\alpha_e}{2\pi} {\rm ln}^2 \frac{Q^2}{q_\perp^2}$ will arise from the final
state soft photon radiation  in higher order calculation.
In particular, at LHC energy, the logarithm terms are sizeable and need to be resummed to all orders to improve
the convergence of the perturbation series. This can be achieved by applying the Collins-Soper-Sterman(CSS)~\cite{Collins:1981uk}
 formalism. The CSS formalism is formulated in the impact parameter space in which
 the large logarithms are resummed into an exponentiation known as the Sudakov factor.
By taking into account the Sudakov factor, the coefficients $\mathcal{A}$ and $\mathcal{C}$
 after the Fourier transform can be rewritten as,
\begin{eqnarray}
\mathcal{A}&=& \frac{(Q^2-2 P_\perp^2)}{P_\perp^2}
x_1x_2\!\int d^2 b_\perp e^{iq_\perp \cdot b_\perp} e^{-S(\mu_b^2,Q^2)}
\\
\nonumber && \times
 \int |k_{1\perp}| J_0(| k_{1\perp}||b_\perp|)  f_1^\gamma(x_1,k_{1\perp}^2) d|k_{1\perp}|
 \int | k_{2\perp}|J_0(| k_{2\perp}||b_\perp|) f_1^\gamma(x_2,k_{2\perp}^2) d| k_{2\perp}|
\\ \nonumber
\mathcal{C} &=&
-2 x_1x_2\! \int  d^2 b_\perp e^{iq_\perp \cdot b_\perp} \cos (4 \theta) e^{-S(\mu_b^2,Q^2)}
\nonumber \\
&& \times
\int |k_{1\perp}| J_2(| k_{1\perp}||b_\perp|)   h_1^{ \perp \gamma}\!(x_1,k_{1\perp}^2) d|k_{1\perp}|
 \int | k_{2\perp}|J_2(| k_{2\perp}||b_\perp|) h_1^{\perp \gamma}\!(x_2,k_{2\perp}^2) d| k_{2\perp}|
\end{eqnarray}
where $\theta$ is the angle between $q_\perp$ and $b_\perp$, and $\mu_b=2e^{-\gamma_E}/|b_\perp|$.
At LHC energy, one can neglect the contributions suppressed by the power of $\frac{m^2}{P_\perp^2}$
in the hard part as shown in the above formulas.
Note that $\cos 2\phi$ asymmetry vanishes at LHC energy under this approximation because it is proportional to
$\frac{m^2}{P_\perp^2}$.  However,  muon mass can not be neglected when computing
 both $\cos 2\phi$ and $\cos 4\phi$ asymmetries at RHIC energy.
At one loop order, the Sudakov factor is given by~\cite{Klein:2018fmp},
\begin{eqnarray}
S(\mu_b^2,Q^2)=
\left \{
\begin{aligned}
&& \frac{\alpha_e}{2\pi} {\rm ln}^2 \frac{Q^2}{\mu_b^2}, \ \ \ \ \ \  \mu_b>m_\mu \\
&& \frac{\alpha_e}{2 \pi} {\rm ln}\frac{Q^2}{m_u^2} \left [
{\rm ln}\frac{Q^2}{\mu_b^2} +{\rm ln} \frac{m_\mu^2}{\mu_b^2}
\right ], \ \ \ \ \ \ \mu_b<m_\mu
\end{aligned}
\right.
\end{eqnarray}
It has been shown that this Sudakov factor plays a crucial
role in correctly reproducing  the high $q_\perp$ tail observed by the ATLAS collaboration~\cite{Klein:2018fmp}.

The distribution of photons coherently generated by the
 charge source inside relativistic nuclei is commonly computed with the Weizs$\ddot{a}$cker-Williams method.
 This quasi-classical method also can be used to determine the linearly polarized photon distribution
 following the similar derivation that relates the dipole  amplitude
 to the various polarized gluon distributions~\cite{Metz:2011wb,Zhou:2013gsa,Boer:2015pni,Boer:2016xqr}.
 Supposing that a nuclei moves along $P^+$ direction, the dominant component of the gauge potential
 is $A^+$ and other components are suppressed by the Lorentz contraction factor $\gamma$.
  Based on this observation,
  after taking  partial integration the photon field strength tensor is approximated as $F_{+\perp}^\mu F_{+\perp}^\nu
 \propto k_\perp^\mu k_\perp^\nu A^+A^+ $, which implies the relation,
$$f_1^\gamma(x,k_{\perp}^2)= h_1^{\perp \gamma} (x,k_{\perp}^2)$$
 In the equivalent photon approximation, one then has~\cite{Bertulani:1987tz,Vidovic:1992ik},
\begin{eqnarray}
xf_1^\gamma(x,k_\perp^2)=xh_1^{\perp \gamma}(x,k_\perp^2)=\frac{Z^2 \alpha_e}{\pi^2} k_\perp^2
\left [ \frac{F(k_\perp^2+x^2M_p^2)}{(k_\perp^2+x^2M_p^2)}\right ]^2
\label{f1h1}
\end{eqnarray}
where $Z$ is the nuclear charge number, and $F$ is the nuclear charge form factor. $M_p$ is proton mass.
 The form factor is often parameterized using  the Woods-Saxon distribution,
\begin{eqnarray}
F(\vec k^2)= \int d^3 r e^{i\vec k\cdot \vec r} \frac{\rho^0}{1+\exp{\left [(r-R_{WS})/d\right ]}}
\end{eqnarray}
where  $R_{WS}$(Au: 6.38fm, pb: 6.62fm) is the radius  and d(Au.:0.535fm, Pb:0.546fm) is the skin depth. $\rho^0$ is the
normalization factor. Alternatively, one can use the  form factor in momentum space
 from the STARlight MC generator~\cite{Klein:2016yzr},
\begin{eqnarray}
F(|\vec k|)=\frac{4\pi \rho^0}{|\vec k|^3 A}\left [ \sin(|\vec k|R_A)-|\vec k|R_A \cos(|\vec k|R_A)\right ]\frac{1}{a^2 \vec k^2+1}
\label{ff}
\end{eqnarray}
where $R_A=1.1 A^{1/3}$fm, and $a=0.7$fm. This parametrization numerically is very close to the Woods-Saxon distribution,
and will be used in our numerical evaluation.
 With all these ingredients, we are ready to perform numerical study of the azimuthal asymmetries in
 lepton pair production in UPCs.

 The numerical results for the computed azimuthal asymmetries in the different kinematical regions are
 presented in Figs.[1-4]. Here the azimuthal asymmetries, i.e. the average value of $\cos (2\phi)$
 and $\cos (4\phi)$ are defined as,
\begin{eqnarray}
\langle \cos(2\phi) \rangle &=&\frac{ \int \frac{d \sigma}{d {\cal P.S.}} \cos (2\phi)  d {\cal P.S.} }
{\int \frac{d \sigma}{d {\cal P.S.}}  \ d {\cal P.S.}}
\\
\langle \cos(4\phi) \rangle &=&\frac{ \int \frac{d \sigma}{d {\cal P.S.}} \cos (4\phi) \ d {\cal P.S.} }
{\int \frac{d \sigma}{d {\cal P.S.}}  d {\cal P.S.}}
\end{eqnarray}
As the $\cos (2\phi)$ azimuthal asymmetry is suppressed by the power of $m^2/P_\perp^2$, it is only sizable
 for di-muon production at RHIC energy.  We plot the $\cos (2\phi)$ asymmetry for muon pair production
at mid-rapidity as the function of the total transverse momentum $q_\perp$ for
three different invariant mass regions at the center mass energy $\sqrt s=200$GeV.
Obviously, the asymmetry decreases with increasing invariant mass as its power behavior indicates. In the lowest invariant mass region
$M_{\mu \mu} \in [0.4, 0.76]$GeV, the asymmetry reaches a maximal value of 10\% percent around $q_\perp=110$MeV.

\begin{figure}[htpb]
\includegraphics[angle=0,scale=1.2]{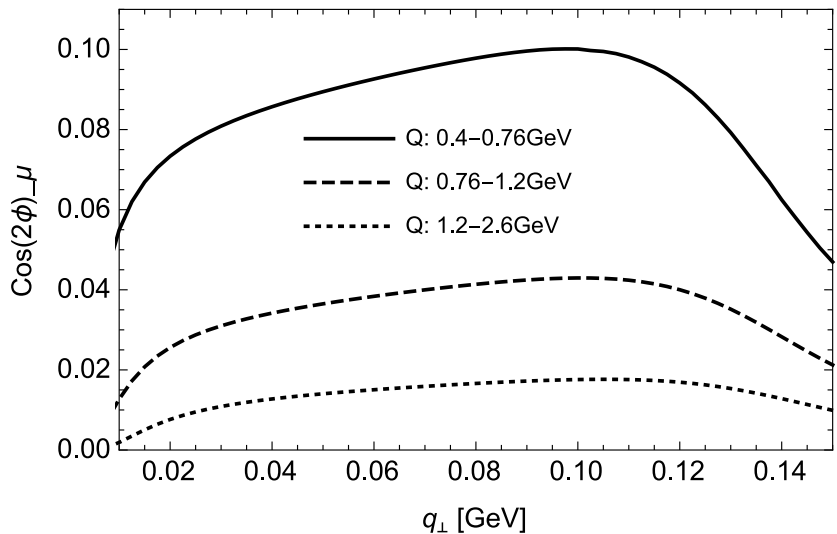}
\caption{ Estimates of the $\cos 2\phi$ asymmetry as the function of $q_\perp$
 for the different  muon pair mass regions 0.4-0.76 GeV, 0.76-1.2 GeV and 1.2-2.6 GeV
 at  $\sqrt {s}=200 $ GeV.
The muon and anti-muon rapidities are integrated over the regions [-1,1].
 } \label{fig1}
\end{figure}

\begin{figure}[htpb]
\includegraphics[angle=0,scale=1.2]{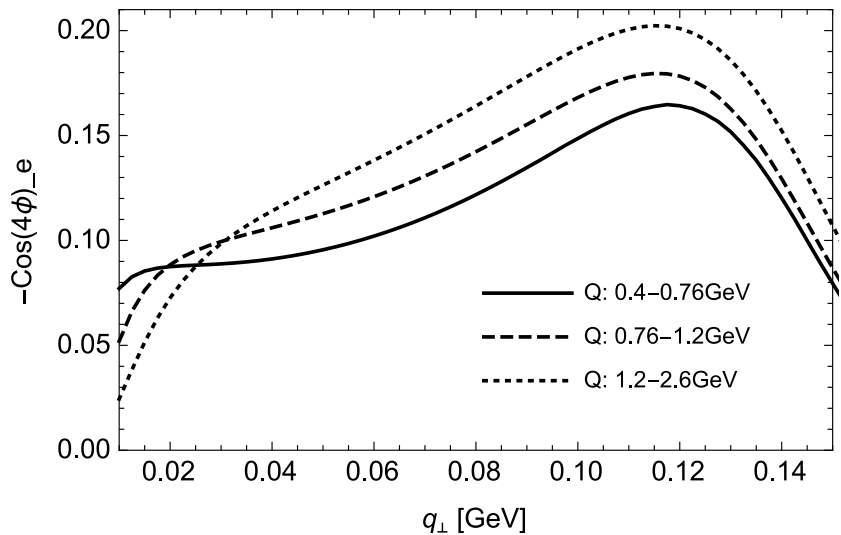}
\caption{ Estimates of the $\cos 4\phi$ asymmetry as the function of $q_\perp$
 for the different  di-electron invariant mass regions 0.4-0.76 GeV, 0.76-1.2 GeV and 1.2-2.6 Gev
 at $\sqrt {s}=200 $ GeV.
The electron and positron rapidities are integrated over the regions [-1,1].
 } \label{fig2}
\end{figure}

\begin{figure}[htpb]
\includegraphics[angle=0,scale=1.2]{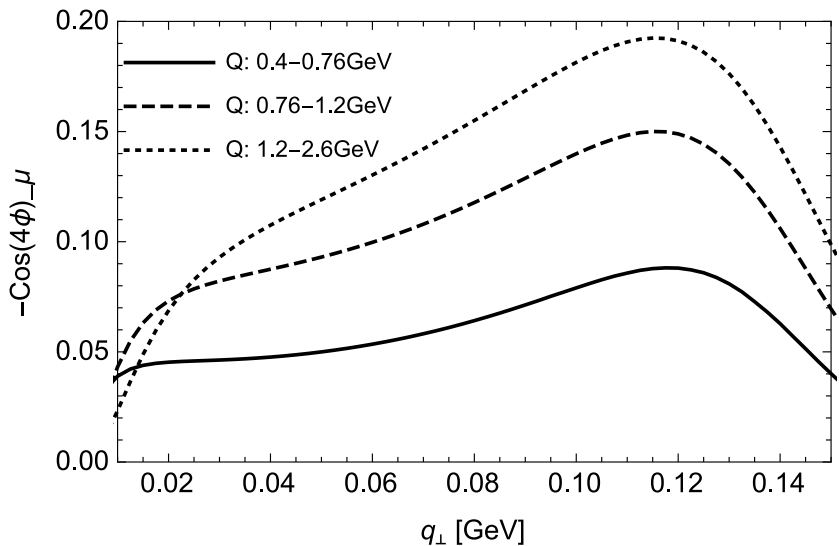}
\caption{ Estimates of the $\cos 4\phi$ asymmetry as the function of $q_\perp$
 for the different di-muon mass regions 0.4-0.76 GeV, 0.76-1.2 GeV and 1.2-2.6 GeV
 at  $\sqrt {s}=200 $ GeV.
The muon and anti-muon rapidities are integrated over the regions [-1,1].
 } \label{fig3}
\end{figure}

For the same kinematical regions at RHIC, we also plot the $\cos 4\phi $ asymmetry for electron pair and muon pair
production. The asymmetry grows with increasing $q_\perp$ until it reaches a maximal value at
 total transverse momentum around 120MeV.  The maximal value of the asymmetry is about 20\% for electron pair production. The $\cos 4\phi $ asymmetry for di-muon production
 is slightly smaller than that for electron pair production in the same kinematical region.
 One sees that  the $\cos 4\phi $ asymmetry drops rather fast  at relatively
 large transverse momentum($>120$MeV).

 The curve for  the $\cos 4\phi $ asymmetry for di-muon production at LHC is presented in Fig.4.
 The  $q_\perp$ dependence of the  asymmetry is similar to these for RHIC energy.
The maximal size of the asymmetry is about 9\% for the invariant mass region [4-45]GeV.
 We further found that
 the Sudakov suppression effect due to final state soft photon radiation reduce the asymmetry
 significantly at relatively large $q_\perp$
  as compared to the lowest order calculation. This may serve as a very clean
 test of the resummation formalism for the QED case.

Note that the terms $\cal B$ and $\cal C$ in Eq.\ref{cs} also affect the di-lepton imbalance  angle distribution.
  The imbalance angle $\delta \phi$ which describes the deviation of the two produced leptons
   from a back-to-back configuration is defined as  $\delta \phi=\phi_1-\phi_2-\pi$ where $\phi_1$ and $\phi_2$
   and the azimuthal angles of the produced leptons' transverse momenta.
 However, at low $q_\perp$, the differential cross section critically depends on the impact parameter~\cite{Vidovic:1992ik}. The main uncertainty of $\delta \phi$ distribution actually comes from the $\cal A$ term in
 Eq.\ref{cs}. To reliably exact the linearly polarized photon distribution via  $\delta \phi$ distribution,
 one has to make a refined analysis by taking into account the impact parameter dependent effect, which
 will be carried out in a future publication.

\begin{figure}[htpb]
\includegraphics[angle=0,scale=1.2]{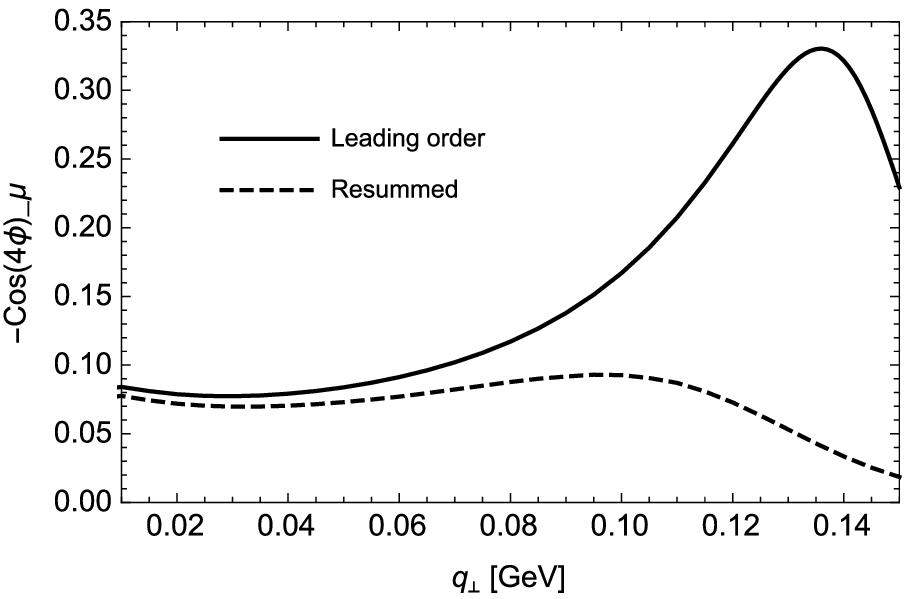}
\caption{ Estimates of the $\cos 4\phi$ asymmetry as the function of $q_\perp$
 for the  di-muon mass region  4-45 GeV
 at  $\sqrt {s}=5.02 $ TeV with and without the resummation effect being incorporated.
 The muon and anti-muon rapidities are integrated over the regions [-1,1].
 } \label{fig3}
\end{figure}

\section{Conclusions}
The unpolarized photon distribution  used to compute physical  observables in ultraperipheral heavy ion collisions
 is commonly determined using the classical external electromagnetic fields of a relativistic charged nuclei.
 Applying this quasi-classical method to the polarized case, one easily finds that the linearly polarized
 photon distribution is identical to the normal unpolarized photon distribution.
 The linearly polarized photon distribution can be  cleanly probed through  the $\cos 2\phi$ and $\cos 4\phi$ azimuthal asymmetries in lepton pair production in
 ultraperipheral heavy ion collisions, where $\phi$ is the angle between lepton pair total transverse momentum
 and individual lepton transverse momentum. We present numerical results for the azimuthal asymmetries
 in  the kinematical regions where the experimental data for di-lepton production has been taken at RHIC
 and LHC. In these kinematical regions,  the magnitudes of the $\cos 4\phi$ azimuthal asymmetry
 for both electron pair and muon pair production are  rather large. And moreover, the  $\cos 2\phi$ azimuthal asymmetry
  in di-muon production at RHIC energy is sizable.
 These findings are very promising concerning a future extraction of $h_1^{\perp\gamma}$ in UPCs
  at RHIC and LHC.
  In our numerical estimation, we also took into account
 the Sudakov suppression effect which reduces the asymmetries significantly at
 relatively large lepton pair transverse momentum.  The Sudakov suppression
 of the azimuthal asymmetry in this process would provide a clean way to test the resummation formalism in the QED case.
 Furthermore, one may expect that this mechanism also plays a role in generating azimuthal asymmetries
 in hadronic heavy-ion collisions.  The study of such initial state effect thus would set a baseline for
 investigating the electromagnetic properties of the quark-gluon plasma created in hadronic heavy-ion collisions
 ~\cite{Kharzeev:2009pj,Asakawa:2010bu,Zha:2018ywo,Klein:2018fmp,Zha:2018tlq,Ye:2018jwq}.

\begin{acknowledgments}
J. Zhou thanks Zhang-bu Xu, James Daniel Brandenburg and Bowen Xiao for helpful discussions.
 J. Zhou has been supported by the National Science Foundations of
China under Grant No.\ 11675093, and by the Thousand Talents Plan
for Young Professionals. Ya-jin Zhou has been supported by the
National Science Foundations of China under Grant
No.\ 11675092.
\end{acknowledgments}

\end{document}